\documentclass[reprint,amsmath,amssymb,aps]{revtex4-2}
\usepackage{graphicx}% Include figure files
\usepackage{dcolumn}% Align table columns on decimal point
\usepackage{bm}% bold math
\usepackage{subfigure}
\usepackage{color}

\begin{document}

\preprint{APS/123-QED}

\title{Spectral Switches of Light in Curved Space} 

\author{Suting Ju$^1$, Chenni Xu$^{1,2}$, and Li-Gang Wang$^1$}
\email{lgwang@zju.edu.cn}
\address{$^1$\colorbox{white}{Zhejiang Province Key Laboratory of Quantum Technology and Device, School of Physics,}\\
\colorbox{white}{Zhejiang University, Hangzhou 310058, Zhejiang, China}
$^2$\colorbox{white}{Department of Physics, The Jack and Pearl Resnick Institute for Advanced Technology, }\\
\colorbox{white}{Bar-Ilan University, Ramat-Gan 5290002, Israel}}
\date{\today}

\begin{abstract}
Acting as analog models of curved spacetime, surfaces of revolution employed for exploring novel optical effects are followed with great interest nowadays to enhance our comprehension of the universe. It is of general interest to understand the spectral effect of light propagating through a long distance in the universe. Here, we address the issue on how curved space affects the phenomenon of spectral switches, a spectral sudden change during propagation caused by a finite size of a light source. 
By using the point spread function of curved space under the paraxial approximation, the expression of the on-axis output spectrum is derived and calculated numerically. 
A theoretical way to find on-axis spectral switches is also derived, which interprets the effect of spatial curvature of surfaces on spectral switches as a modification of effective Fresnel number. 
We find that the spectral switches on surfaces with positive Gaussian curvature are closer to the source, 
compared with the flat surface case, while the effect is opposite on surfaces with negative Gaussian curvature. 
We also find that the spectral switches farther away from the light source are more sensitive to the change in Gaussian curvature. 
This work deepens our understanding of the properties of fully and partially coherent lights propagating on two-dimensional curved space. 
\end{abstract}

\maketitle

\section{Introduction}
Phenomena in curved spacetime and the physics behind them have been fascinating people since Einstein proposed General Relativity. 
For example, black holes and their event horizons around them, gravity waves, the universe itself, and so on, all are still mysterious. 
In most cases, only passive observation is possible as celestial bodies are far away and their gravitational effect is weak. 
Fortunately, the idea of analogy provides a new method for us to research them. 
Nowadays a variety of analog models have been developed to investigate effects in curved spacetime, 
such as the water tank \cite{ref1-1-1}, the BEC system \cite{ref1-1-2,ref1-1-3}, slow light in fiber optics \cite{ref1-1-4}, and graphene \cite{ref1-1-5}. One can also refer to these reviews \cite{ref1-1,ref1-2}. 

In condensed matter physics, the theory of quantum particles confined in curved space came up and the geometric potential was discussed \cite{ref2-1}. 
Since the paraxial Helmholtz equation of light on a surface is very similar to Schr\"{o}dinger's equation, 
light and plasmonic beams are used to study the propagation mechanism of quantum particles and quantum effects \cite{ref2-2,ref2-3,ref2-4,ref2-5,ref2-5-1}. 
Batz and Peschel then considered optical effects in curved space and regarded it as an analog model of General Relativity \cite{ref2-6}. 
Unlike the analog models mentioned above, the curved spacetime was reduced in dimensions and turned into a two-dimensional surface in this case. 
Such models can visually display the geometry of curved space and probably provide new ideas for manufacturing optical devices. 
There have been some interesting researches about general relativistic phenomena and optical phenomena in curved space \cite{ref2-6,ref2-7,ref2-8,ref2-9,ref2-10,ref2-11,ref2-12,ref2-13,ref2-14}. 
Among them, surfaces with constant Gaussian curvature can act as analog models of universes with non-vanishing cosmological constants, and are thus of particular interest in various works  \cite{ref2-7,ref2-8,ref2-9,ref2-10,ref2-11}. 
For example, Gaussian beams can self-focus on this type of surfaces with positive Gaussian curvature  \cite{ref2-7}. It is theoretically and experimentally found that light doesn’t always gain coherence on this kind of surface, as stated on Hanbury Brown and Twiss effect \cite{ref2-8}. 
The Wolf effect, which is caused by the correlation of a light source, is found to be enhanced by positive curvature and suppressed by negative curvature \cite{ref2-9}. 
The shape-preserving accelerating beams are theoretically and experimentally realized on the sphere \cite{ref2-10}. 
Recently, the Branched flows of light are also engrossingly observed on a semi spherical bubble \cite{ref2-11}. 
Besides the surfaces of constant Gaussian curvature, Flamm’s paraboloid is also adopted because it is derived from the Schwarzschild metric. 
Two Flamm’s paraboloids can be joined together and form an analog model of wormholes. 
The effect of tunneling was found when wave packets propagate in this model \cite{ref2-12}. 
There are also some theoretical works considering the propagation of light on more general surfaces of revolution \cite{ref2-13,ref2-14}. 

Here, we are going to consider the effect of spectral switches of light, a kind of spectral sudden change, on surfaces of revolution. 
Spectral change of light has attracted a lot of attention, 
because it can offer or influence the information we get from the spectrum. 
For example, the Doppler effect can tell us the relative velocity between a source and an observer. 
Its further work is the angular Doppler effect. 
It has applications in rotational Raman scattering, fluorescence doublets, and so on \cite{ref3-1}. 
The theory was also developed to detect spinning objects in astronomy \cite{ref3-2}. Another example is the Wolf effect \cite{ref3-3,ref3-4}. 
It indicates that the spectrum of some partially coherent sources would suffer a spectral shift even though the light propagates in free space. 
This effect has been experimentally approved in different systems \cite{ref3-5,ref3-6}, such as the partially coherent light sources \cite{ref3-5} and the acoustic experiment \cite{ref3-6}. 
An application of the Wolf effect is spatial-coherence spectral filters, 
which can be applied in separation of neighboring spectral lines, processing optical signals, and optical coding \cite{ref3-7,ref3-8}.

Exploration of the spectral shift induced by a confined aperture in the near zone led to the revelation of the spectral switch, as initially proposed by \cite{ref4-1}. 
And the spectral switches was found to be closely related to spectral anomalies \cite{ref4-2}. 
The meaning of the spectral switch is the rapid spectral change between redshift and blueshift at certain specific regions. At the positions of occurring the spectral switch, the spectrum changes drastically and possesses two peaks with the same height. This effect could be induced by the diffraction of fully coherent sources or partially coherent sources obeying the scaling law \cite{ref4-2}. For partially coherent sources violating the scaling law, the spectral switch is both diffraction-induced and correlation-induced \cite{ref4-3}. The spectral switch has been experimentally observed \cite{ref4-4,ref4-5,ref4-6,ref4-7}, including circular and rectangular apertures \cite{ref4-4}, 
far-field off-axis spectral switches \cite{ref4-5}, $1\times N$ spectral switch \cite{ref4-6}, and spectral switches in Young's double-slit experiment \cite{ref4-7}. 
There are also researches about spectral switches of scattering spectra \cite{ref4-8,ref4-9,ref4-10}. The effect of spectral switches has its important application in lattice spectroscopy \cite{ref4-11}, spatial-coherence spectroscopy \cite{ref4-12}, and digital data transmission \cite{ref4-13,ref4-14}. Although so many works have been done, optical systems in flat space were focused by all the previous researches. In this work, we are going to  explore whether spectral switches can happen on two-dimensional curved space and the way the spatial curvature influences the positions of spectral switches.

\section{The output spectrum of a finite light source in curved space}

\subsection{The point spread function in curved space}

\begin{figure*}[t]
	\centering
	\subfigure[]{\includegraphics[scale=0.35]{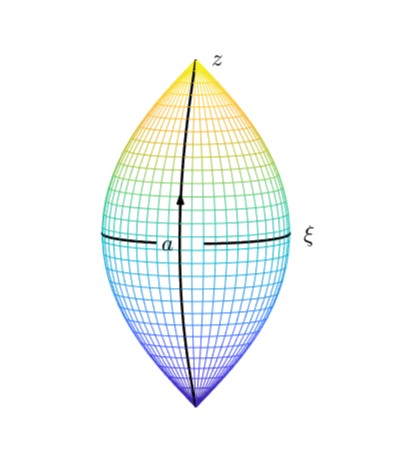}\label{fig:subfigSpindle}}
	\subfigure[]{\includegraphics[scale=0.35]{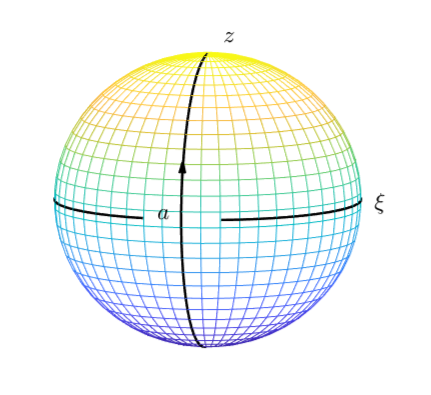}\label{fig:subfigSphere}}
	\subfigure[]{\includegraphics[scale=0.35]{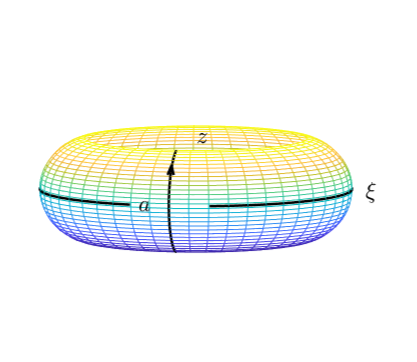}\label{fig:subfigBulge}}
	\subfigure[]{\includegraphics[scale=0.35]{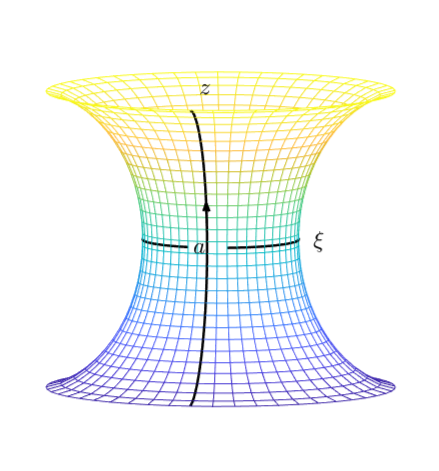}\label{fig:subfigHyperboloid}}
	\caption{Different types of surfaces with constant Gaussian curvature and the curvilinear coordinates thereon. 
The two parameters of the surface are $z$ and $\xi$. 
The slit is settled on $(0,0)$ and its length is $2a$. 
Light propagates from the slit along the $z>0$ part of the longitude. 
(a) to (c) are surfaces with constant positive Gaussian curvature 
and (d) is a surface with constant negative Gaussian curvature. 
When $K$ is positive, the relative size between $R$ and $r_{0}$ can influence the shape of the surface, such as 
(a) a spindle with $R>r_{0}$, (b) a sphere with $R=r_{0}$ and (c) a bulge with $R<r_{0}$. 
(d) is always the shape of a hyperboloid no matter how we choose $R$ and $r_{0}$. 
}
	\label{fig:plotSurface}
\end{figure*}

A surface of revolution is produced by the rotation of a curve about a certain axis, and the curve is 
called the generatrix. Each point on this surface can be expressed as $\vec{s}=(r(z)\cos\varphi,r(z)\sin\varphi,h(z))$
where $r(z)$ is the expression of the generatrix and satisfies $(\mathrm{d} r/\mathrm{d} z)^{2}+(\mathrm{d} h/\mathrm{d} z)^{2}=1$. 
The parameter $z$ represents the proper length along the generatrix and $\varphi$ represents the rotation angle. 
The Gaussian (intrinsic) curvature $K=1/R_{1}R_{2}$ and the extrinsic (mean) curvature $H=(1/R_{1}+1/R_{2})/2$ 
are often used to describe the characteristics of a surface, where $R_{1}$ and $R_{2}$, respectively, represent the radii of the maximal and minimal circles that are tangent to the surface at the same point. If these two circles are on the opposite sides of the surface, the Gaussian curvature $K$ will be negative. When the surface has a constant Gaussian curvature, then its generatrix is
\begin{equation}
r(z)=r_{0}\cos_{q}(z/R)=\left\{\begin{matrix}
r_{0}\cos(z/R) & q=1\\
r_{0}\cosh(z/R) & q=-1
\end{matrix}\right.
\end{equation}
where $q=sgn(K)$, $r_{0}$ is the radius of the equator. The parameter $R$ is the radius of the surface's curvature and then its Gaussian curvature is $\left | K \right | =1/R^{2}$. In the situation of $K>0$, the relative difference between $R$ and $r_{0}$ can influence the shape of the surface. 
For $R>r_{0}$, it is a spindle with $\left | z \right |<\pi R/2$; for $R=r_{0}$, it becomes a sphere with 
the same range value for $z$; while for $R<r_{0}$, it changes into a bulge with $\left | z \right |<R\sin^{-1}(R/r_{0})$. When $K<0$, the shape of the surface is always hyperboloid with $\left | z \right |<R\sinh^{-1}(R/r_{0})$. The shape of the surfaces with different $R$ and $r_{0}$ 
are shown in Fig.~\ref{fig:plotSurface}.

On a general surface of revolution, the light propagation can be described by the following scalar Helmholtz equation \cite{ref2-6}
\begin{equation}
(\bigtriangleup_{g}+k^{2})\Psi=-(H^{2}-K)\Psi,
\label{eq:Helmholtz}
\end{equation}
where $\Psi$ is the light field, 
$\bigtriangleup_{g}=\frac{1}{\sqrt{g}}\partial_{i}(\sqrt{g}g^{ij}\partial_{j})$ is the covariant Laplace operator on a surface with the two surface coordinate parameters $x^{i}$ and $x^{j}$, 
$g^{ij}$ is the inverse form of the metric of the surface $g_{ij}$, 
$g$ is the determinant of metric $\det(g_{ij})$, 
$k=\omega/c$ is the wave number, and $c$ is the speed of light in vaccum. 
With $z$ for the proper length along the longitudinal direction and $\xi=r_{0}\varphi$ for the arc length on the equator, the metric of a general surface of revolution can be written as 
\begin{equation}
\mathrm{d}s^{2}=\mathrm{d}z^{2}+(\frac{r(z)}{r_{0}})^{2}\mathrm{d}\xi ^{2}.
\label{eq:metric}
\end{equation}
Substituting Eq.~(\ref{eq:metric}) into Eq.~(\ref{eq:Helmholtz}) and assuming $\Psi=\frac{A}{\sqrt{r(z)}}v(z,\xi)e^{ikz}e^{i\phi}$, under the paraxial approximation, one can obtain 
\begin{equation}
2ik\frac{\partial v}{\partial \Xi}+\frac{\partial^{2} v}{\partial \xi^{2}}=0,
\label{eq:Schrodinger}
\end{equation}
where $\Xi(z)=\int_{0}^{z} (\frac{r_{0}}{r(z')})^{2}dz'$ is the effective propagation distance which has elaborated in \cite{ref2-13} 
and $\phi=\frac{1}{2k}\int_{0}^{z}V_{eff}(z')dz'$ is the additional phase caused by curved surfaces 
with $V_{eff}(z)=\frac{1}{4}(\frac{r'(z)}{r(z)})^{2}-\frac{1}{2}\frac{r''(z)}{r(z)}$. 
Equation ~(\ref{eq:Schrodinger}) has a similar form as the Schr\"{o}dinger equation. 
On the right side of Eq.~(\ref{eq:Helmholtz}), the extrinsic curvature $H$ and the intrinsic curvature $K$ have been neglected since the radius of surface $R$ we considered here is much larger than the wavelength. 
For example, the corresponding values of $K$ and $H^{2}$ in table experiments are usually the order of $10^{0}\sim 10^{2}$ $m^{-2}$ while $k^2$ is roughly $10^{14}$ $m^{-2}$. 
By solving Eq.~(\ref{eq:Schrodinger}), one has obtained the point spread function of light that spreads along $\xi=const$ (the line of longitude) on the surface, which is \cite{ref2-13}
\begin{equation}
h_{q}(\eta,\xi,z)=\sqrt{\frac{kr_{0}}{2\pi i \Xi r(z)}}e^{\frac{ik}{2\Xi}(\eta-\xi)^{2}}e^{ikz}e^{i\phi},
\label{eq:psf}
\end{equation}
where $\eta$ is the abscissa on the line of non-equator latitude. 

\subsection{On-axis spectrum of light in curved space}

In order to consider a finite light source, here we assume that the size of the source is $2a$, which can also be seen as that a beam of polychromatic partially coherent light passes through a slit with width of $2a$, as shown in Fig.~\ref{fig:plotSurface}. The light propagates along the direction of $z>0$ (i.e., the line of longitude on the surfaces of revolution).
The center of the slit is located at the origin of the coordinate $(z=0,\xi=0)$. 
Our aim is to study the spectral change along the propagation, 
so we first need to obtain the on-axis output spectrum. 

For a polychromatic partially coherent plane light, its cross-spectral density at the initial place can be expressed as \cite{ref4-1}
\begin{equation}
W_{in}(\xi_{1},\xi_{2},z=0,\omega)=S_{0}(\omega)e^{-\frac{(\xi_{2}-\xi_{1})^{2}}{2\sigma(\omega)^{2}}},
\label{eq:W_in}
\end{equation}
where $S_{0}(\omega)$ is the initial spectrum, and $\sigma(\omega)$ is the rms correlation width. Without loss of generality, we assume the source satisfies the Scaling law and let $\sigma(\omega)=\sigma_{0}\frac{\omega_{0}}{\omega}$. 
The parameters $\xi_{1}$ and $\xi_{2}$ represent different points at the initial plane of the light source. We also assume that the spectrum of the source is a Lorentz type with $\omega_{0}$ the center (or peak) frequency and $\Gamma$ the half-width of spectral line which is expressed by
\begin{equation}
S_{0}(\omega)=\frac{\Gamma^{2}}{(\omega-\omega_{0})^{2}+\Gamma^{2}}.
\end{equation}
According to the theory of partially coherent light \cite{MandelWolf1995}, the cross-spectral density of the light field at the output plane $z>0$ can be written as
\begin{equation}
W_{out}(\eta_{1},\eta_{2},z,\omega)=\int\limits_{-a}^{a} \int\limits_{-a}^{a} W_{in}h_{q}(\eta_{1},\xi_{1},z)h_{q}^{*}(\eta_{2},\xi_{2},z)d\xi_{1}d\xi_{2},
\label{eq:Wout}
\end{equation}
where $(\eta_{1},z)$ and $(\eta_{2},z)$ are the different points on the output end, $h_{q}$ is the point spread function introduced above and $h_{q}^{*}$ is the conjugate function of $h_{q}$. 

Now we concentrate our aim on the evolution of the light spectrum along the propagation axis (i.e., the line of the longitude by choosing $\eta_{1}=\eta_{2}=0$) to achieve the on-axis 
output spectrum, which can be expressed as
\begin{align}\nonumber
S(z,\omega)&=W_{out}(0,0,z,\omega)\\
&=\frac{\omega a^{2}S_{0}(\omega)r_{0}}{c\pi\Xi r(z)}\int\limits_{0}^{1}\int\limits_{0}^{1}e^{-\frac{a^{2}(u_{1}-u_{2})^{2}}{2\sigma(\omega)^{2}}}e^{i\frac{\omega a^{2}}{2c\Xi}(u_{1}^{2}-u_{2}^{2})}du_{1}du_{2}.
\label{eq:Sout}
\end{align}
In the case of fully coherent light, that is to say when $\sigma_{0} \to \infty$, the output spectrum becomes
\begin{equation}
S(z,\omega)=\frac{r_{0}}{r(z)}S_{0}(\omega)[C(t_{\Xi})^{2}+S(t_{\Xi})^{2}],
\label{eq:FullyCoherent}
\end{equation}
where $C(x)=\int_{0}^{x}\cos(\pi u^{2}/2)du $ is the cosine Fresnel integral, $S(x)=\int_{0}^{x}\sin(\pi u^{2}/2)du $ is the sine Fresnel integral, and here $t_{\Xi}=\sqrt{\frac{\omega a^{2}}{c\pi\Xi}}$. Note that ${\Xi}$ is a function of $z$. 
For the sake of simplicity, we define $z_{0}=a^{2}/\lambda_{0}=\omega_{0}a^{2}/2\pi c$ \cite{ref4-1} and use $z/z_{0}$ to describe thee propagation distance on axis. 
The Fresnel number at $z_{0}$ is $1$, so $z/z_{0}<<1$ means the near-field region and $z/z_{0}>>1$ means the far-field region. 

\section{Results and Discussions}

\begin{figure*}[t]
\includegraphics[width=1\textwidth]{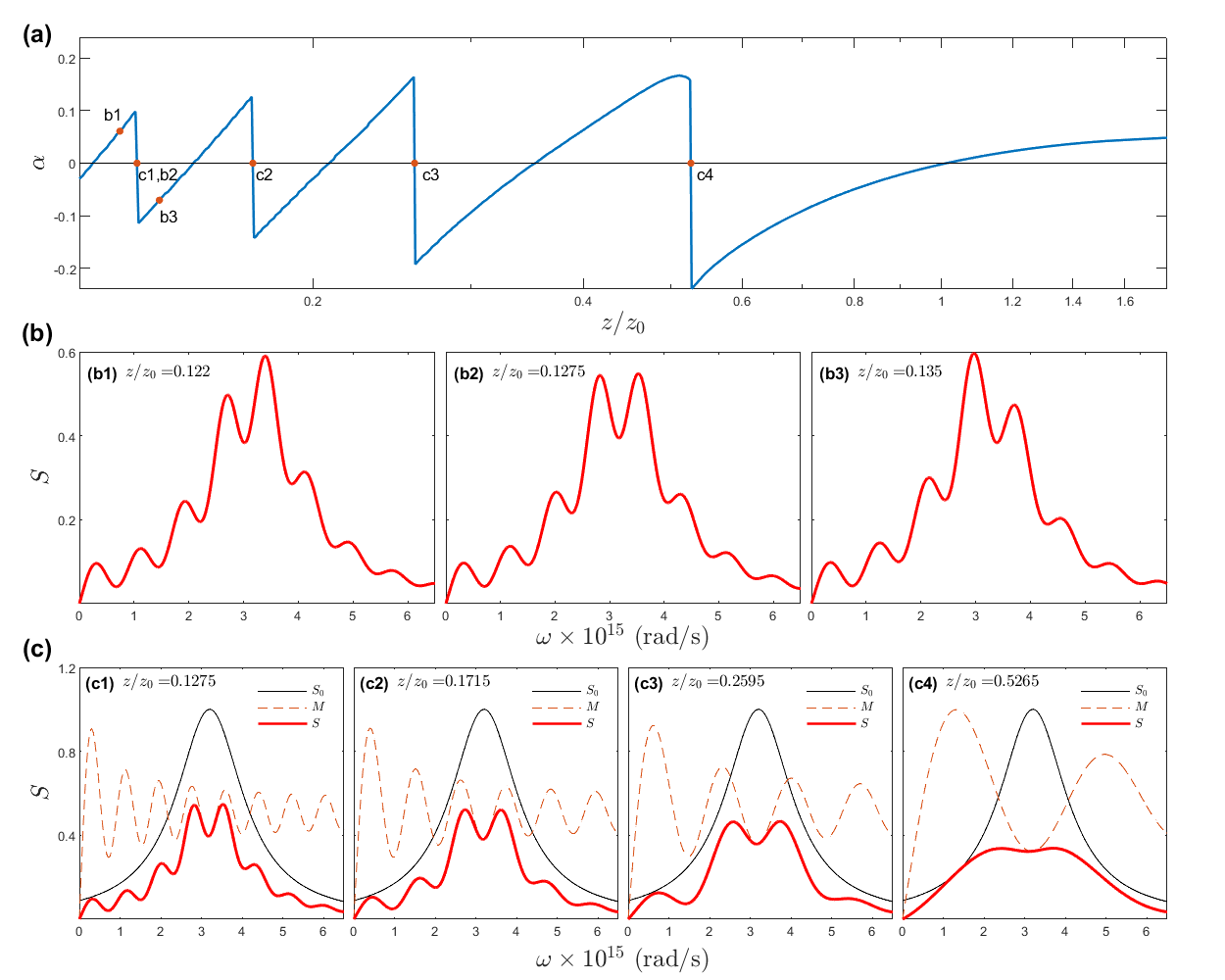}
\caption{\label{fig:ExplainSpectralSwitch}
Effect of spectral switches of a fully-coherent light source on the surface with the constant Gaussian curvature $K=4$ m$^{-2}$.
(a) The relative spectral shift curve along the propagation axis. The discontinuity points occur at $z/z_{0}=0.1275,0.1715,0.2595,0.5265$ standing for the positions of spectral switches. 
(b) Spectral distributions of light at different on-axis positions $z/z0=0.1220$ (b1),$0.1275$ (b2), and $0.1350$ (b3) near one of spectral switches ($z/z0=0.1275$).
(c) Spectrum profiles at different locations of occurring spectral switches, which are denoted by the symbols of c1, c2, c3, and c4, corresponding to the 4th, 3rd, 2nd, and 1st spectral switches caused by different valleys of the modifier. Note that in (c1-c4), the thin black curves are the initial spectra, the red thick curves are the output spectra, and the red dashed curves denote the modifiers at different propagation distances. Here, the size of the slit is $2a=1.0$ mm, and the parameters of the initial spectrum are the center frequency $\omega_{0}=3.2\times10^{15}$ rad/s and the half-width $\Gamma=1\times10^{15}$ rad/s. 
}
\end{figure*}

For the sake of discussing the red- or blue-shift of a spectral line, here we define a relative spectral shift that is the relative difference between the peak frequency of the output spectrum at certain position and the peak frequency of the initial spectrum, which can be expressed by
\begin{equation}
\alpha=\frac{\omega_{1}-\omega_{0}}{\omega_{0}},
\end{equation}
where $\omega_{1}$ is the corresponding peak frequency of the output spectrum. From Eq.~(\ref{eq:Sout}), one can find every $\omega_{1}$ numerically from the on-axis output spectral information. When there are discontinuity points in the change curve of $\omega_{1}$ vs the propagation distance $z$, we call such effect as spectral switches. 

Fig.~\ref{fig:ExplainSpectralSwitch} shows the effect of spectral switches on a surface with positive constant Gaussian curvature. 
From Fig.~\ref{fig:ExplainSpectralSwitch}(a), we observe the change of the relative spectral shift $\alpha$ along the propagation axis. It is seen that the relative spectral shift $\alpha$ increases from the red to blue shifts repeatedly as the propagation distance increases and it suffers the sudden drops from the blue shift to red shift at certain positions (denoted by the points c1, c2, c3,c4) where spectral switch happens.  
As is seen in Fig.~\ref{fig:ExplainSpectralSwitch}(b), it shows the changes of spectral distributions near one of  spectral switches, and the ripples of spectral distributions originate from the diffraction of light by the slit. Due to the diffraction of light, the maximal spectral intensity gradually shifts and redistributes with the increase of propagation distance $z$, and at specific positions there are two maximal (equal) spectral intensities indicating the occurrence of the spectral switch. Accordingly, the relative spectral shift $\alpha$ changes drastically from positive to negative. 
Spectral switches occur on these critical points and their spectra are shown in Fig.~\ref{fig:ExplainSpectralSwitch}(c). 

Furthermore, we figure out a theoretical method to search for these critical positions of spectral switches, 
which can help us understand this effect on curved surfaces. 
The output on-axis spectrum can be rewritten as
\begin{equation}
S(z,\omega)=S_{0}(\omega)M(z,\omega),
\label{eq:modifer}
\end{equation}
where the function $M(z,\omega)$ is usually recognized as the modifier (or the transfer function). 
One needs to find positions where the output spectrum has two peaks with equal heights and a valley between them, 
so one can turn to search for the valleys of the modifier $M(z,\omega)$. 
From Fig.~\ref{fig:ExplainSpectralSwitch}(c) we can find that the modifier has many peaks and valleys and 
different spectral switches occur when different valleys of the modifier correspond to the central frequency $\omega_{0}$. 
To facilitate further discussion, we classify the spectral switches based on the order of the valley that directly causes the spectral switch. 
According to the classification, the four spectral switches shown in Fig.~\ref{fig:ExplainSpectralSwitch}(c) are 
the 4th spectral switch, the 3rd spectral switch, the 2nd spectral switch, and the 1st spectral switch respectively from left to right, 
that is, spectral switches with larger orders are closer to the source. 
Actually there are more spectral switches closer to the source, 
but we only show the four farthest spectral switches in Fig.~\ref{fig:ExplainSpectralSwitch}(c) for simplicity. 

\begin{figure}[b]
\includegraphics[width=0.48\textwidth]{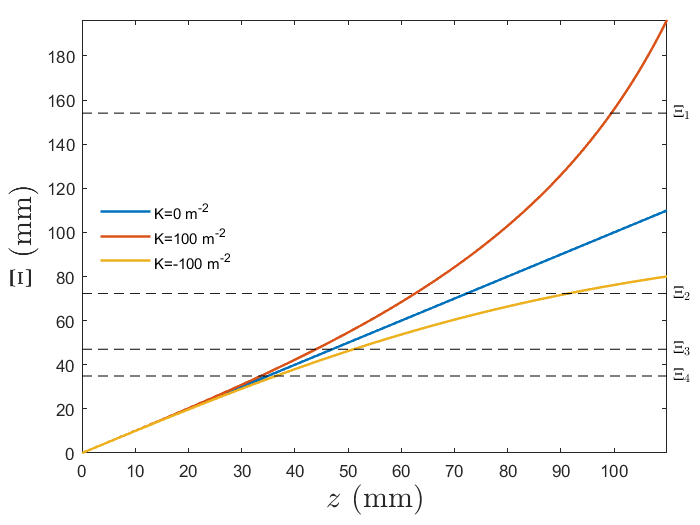}
\caption{\label{fig:Xi}Effective propagation distance $\Xi$ changing with $z$ 
on a surface with positive constant curvature of $K=100$ 
 m$^{-2}$, a flat surface, and a surface with negative constant curvature of $K=-100$ m$^{-2}$. 
Black dashed lines denote the values of the effective propagation distance $\Xi_{m}$ satisfied for different orders $m$ of spectral switches. The intersection points where the black dashed lines cross the curves of effective propagation distances represent the solutions to the theoretical approximation of spectral switches.}
\end{figure}

When the source is fully coherent with $\sigma_{0}/a\to\infty$, 
the modifier can be written as $M(z,\omega)=r_{0}/r(z)[C(t_{\Xi})^{2}+S(t_{\Xi})^{2}]$ according to Eq.~(\ref{eq:FullyCoherent}) 
where $t_{\Xi}=\sqrt{\omega a^{2}/\pi c\Xi}$ can be considered as a function of $\omega$ in this case 
and the modifier $M(z,\omega)$ can be seen as $M(z,t_{\Xi}(\omega))$. 
By taking the derivative of the modifier to get the equation of extreme point, we have 
\begin{equation}
\frac{\partial M}{\partial \omega}=\frac{\partial M}{\partial t_{\Xi}}\frac{\partial t_{\Xi}}{\partial \omega}=\frac{\partial M}{\partial t_{\Xi}}\sqrt{\frac{a^{2}}{4\pi c\omega\Xi}}=0. 
\end{equation}
Since there is always $\sqrt{a^{2}/4\pi c\omega\Xi}>0$, our aim is then turned to search for the zero points 
in the expression
\begin{equation}
\partial M/\partial t_{\Xi}=C'(t_{\Xi})C(t_{\Xi})+S'(t_{\Xi})S(t_{\Xi})=0, 
\label{eq:func}
\end{equation}
and find the local minimum points in $M(z,t_{\Xi})$. 
The zero points in Eq.~(\ref{eq:func}) can be obtained numerically and 
we denote the value of the independent variable of these zero points as $v_{i},(i=0,1,2,3,...)$. 
The local minimum points occur where $i$ is an even positive integer. 
Therefore, the condition for the $m$-th local minimum of the modifier $M(z,\omega)$ to occur is 
$t_{\Xi}=v_{2m},(m=1,2,3,...)$. 
When the $m$-th local minimum point of the modifier coincides with the central frequency $\omega_{0}$, 
the on-axis output spectrum will be very close to the spectral switch, 
so the approximation expression of the $m$-th on-axis spectral switch is 
\begin{equation}
\Xi_{m}=\frac{\omega_{0} a^{2}}{\pi cv_{2m}^{2}},
\label{eq:positiontheory}
\end{equation}
with $m=1,2,3,...$.  
We can understand Eq.~(\ref{eq:positiontheory}) in another way that 
when the effective propagation distance $\Xi$ reaches some specific values, 
the spectral switches will occur. For example, the expression of effective propagation distance on surfaces with constant Gaussian curvature 
is $\Xi$=$R\tan(z/R)$ for positive curvature and $\Xi$=$R\tanh(z/R)$ for negative curvature. 
The on-axis positions for generating identical spectral switches at the same effective propagation distances are depicted in Fig.~\ref{fig:Xi}. These positions correspond to varying values of $z$ on surfaces of distinct curvatures.

Inspired by the previous work \cite{ref4-2}, spectral switches in 3-dimensional flat space occur 
when the Fresnel number at the center frequency is an even integer, that is $N(\omega_{0})=2m\phantom{1}(m=1,2,3...)$.
If we define the effective Fresnel number $N_{eff}=\omega a^{2}/2\pi c\Xi$ on surfaces of revolution, 
the quantity $t_{\Xi}(\omega_{0})$ can be expressed as $t_{\Xi}=\sqrt{2N_{eff}}$. 
Analogously, spectral switches on 2-dimensional surfaces of revolution occur 
when the effective Fresnel number at the center frequency satisfies $N_{eff}(\omega_{0})=v_{2m}^{2}/2,(m=1,2,3...)$. 
Hence, spectral switches for a fully coherent source are also diffraction-induced. 
The difference comes from the infinitesimal $dr$ in 2D space and $rdrd\theta$ in 3D space.
In 2D curved space, another notable distinction arises from the non-uniform values of local maxima in the modifier. Consequently, the assigned position corresponds not to the precise spectral switch point but rather to a red-shift point slightly preceding it. In contrast, within 3D flat space, the derived expression enables the determination of the exact spectral switch points. The specific form of the modifier in 3D flat space can be referenced in Ref.\cite{ref4-2}. 

\begin{table}[h]
\caption{\label{tab:table1}
Comparison between the positions of on-axis spectral switches obtained from Eq.~(\ref{eq:positiontheory}) and the exact numerical method. The four on-axis spectral switches are chosen in the surface of revolution with the Gaussian curvature $K=4$ m$^{-2}$ and the light source is fully coherent.}
\begin{ruledtabular}
\begin{tabular}{ccccc}
\textrm{Orders}&
\textrm{1st}&
\textrm{2nd}&
\textrm{3rd}&
\textrm{4th}\\
\colrule
Eq.~(\ref{eq:positiontheory}) ($z/z_{0}$) & 0.5310 & 0.2628 & 0.1727 & 0.1287\\
Numerical Method ($z/z_{0}$) & 0.5265 & 0.2595 & 0.1715 & 0.1275\\
Relative error & 0.75\% & 1.07\% & 0.99\% & 0.54\%\\
\end{tabular}
\end{ruledtabular}
\end{table}

Table 1 presents the comparison of the predicted positions of on-axis spectral switches between the exact numerical method and the theoretical Eq.~(\ref{eq:positiontheory}).  
We use the numerical method to estimate the effectiveness of Eq.~(\ref{eq:positiontheory}). As is shown in Table~\ref{tab:table1}, we choose four different spectral switches, calculate their positions in both ways and compare the results. The spectral-switch positions given by Eq.~(\ref{eq:positiontheory}) are all slightly larger than their exact positions by the numerical method. The decrease in the values of the local maxima of the modifier with increasing frequency, as illustrated in Fig. 2(c), results in a subtle advanced shift in the positions where spectral switches occur. However, the relative errors are all under $2\%$, which shows that Eq.~(\ref{eq:positiontheory}) works for the cases of fully coherent light. 
Hence, it is affirmed that Eq.~(\ref{eq:positiontheory}) can be initially employed to approximate the positions, followed by a numerical method to precisely determine the locations of spectral switches.

\begin{figure}[h]
\includegraphics[width=0.48\textwidth]{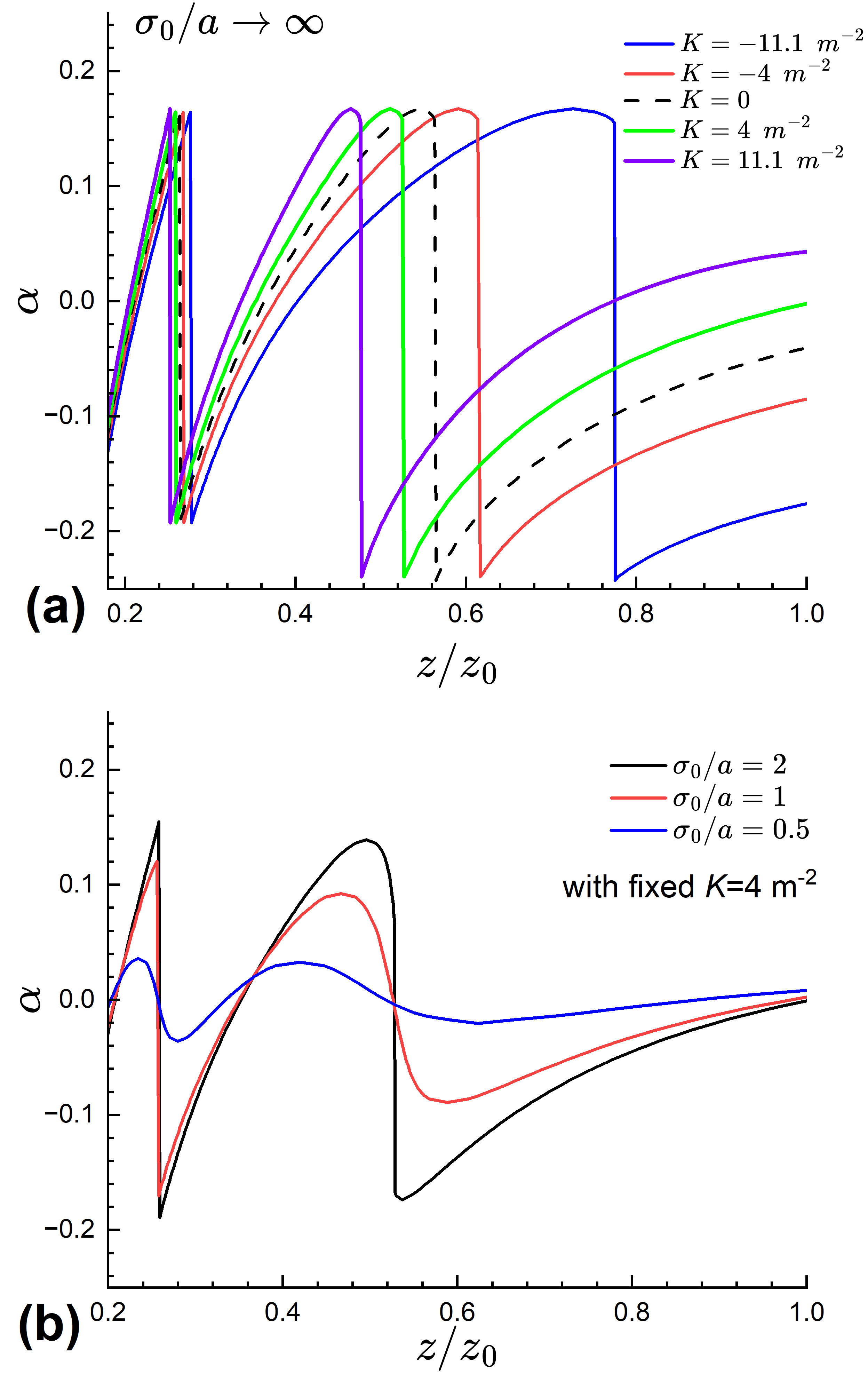}
\caption{Effects of (a) the curvatures of surfaces and (b) the spatial correlation of light on the behaviors of spectral switches. In (a), the light sources are fully coherent (i.e., $\sigma_{0}/a \to \infty$). In (b), the surface has the constant Gaussian curvature $K=4$ m$^{-2}$.}
\label{fig:4fig}
\end{figure}

Now let us turn to discuss how the curvatures of surfaces and the spatial correlation of the light source 
influence the behaviors of spectral switches. 
In Fig.~\ref{fig:4fig}(a), it displays that the curvatures of curved surfaces do not impact the redshift or blueshift values on either side of the discontinuity point (the location of spectral switch). However, the curvatures do play a role in influencing the rates of change for the relative spectral shifts and the positions of spectral switches. It is seen that, for the $K<0$ case, the relative spectral shift curve is stretched larger as the absolute value of $K$ increases. This property indicates that the spectral shift is decelerated in the curved space with negative constant curvatures, resulting in the locations of occurring spectral switches farther away from the light source.  
In the $K>0$ case, the curve of the relative spectral shift is shrunk, indicating the acceleration of the spectral shift in curved space with positive Gaussian curvatures and resulting in the locations of spectral switches closer to the source for larger values of $K$. With the increase of $K$, these curves are more and more shrunk. 

In Fig. 4(b), it shows the changes of the relative spectral shifts under different values of the spatial correlation lengths $\sigma_{0}$. When the light sources are partially coherent, the magnitudes of the relative spectral shift become smaller as the value of $\sigma_{0}$ decreases. Meanwhile, as the value of $\sigma_{0}$ decreases, the relative spectral shift curve smooths away the discontinuity point at the farthest position of the light source, and the amount of blue shift and redshift corresponding to other discontinuity points will also decrease.
For example, the spectral switch happens at nearby $z/z_{0}=0.26$ on the flat surface. Its blueshift has 
decreased from $0.167$ to $0.123$ and its redshift value has decreased from $0.193$ to $0.171$. Thus, as the source changes from fully coherent light to partially coherent light, the valley of the modifier becomes shallower, which may lead to the valley of the output spectrum also becoming shallow and disappearing. That may finally result in the disappearance of the spectral switch phenomenon.

\begin{figure}[t]
\includegraphics[width=0.48\textwidth]{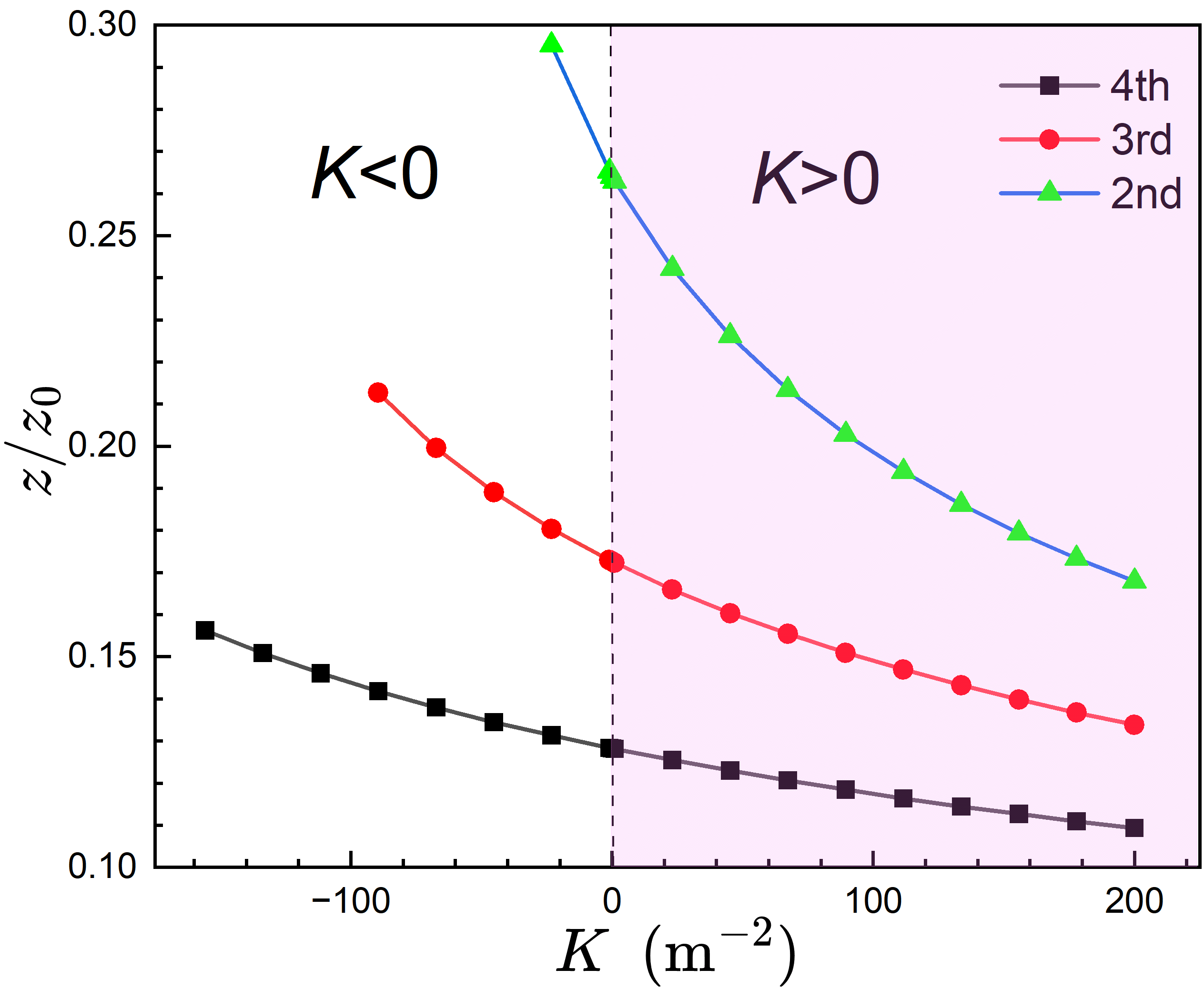}
\caption{\label{fig:Figure5}
Effect of space curvatures on positions of spectral switches. 
Here we choose the three kinds of spectral switches from top to bottom corresponding to the 2nd-, 3rd-, and 4th-order spectral switches.}
\end{figure}

In order to study the change of the fully coherent light source's spectral switch position with Gaussian curvature $K$ in more detail, 
we expand the range of $K$ to $[-200,200]$ m$^{-2}$, and select 21 different $K$ in this range, including 10 negative $K$, 10 positive $K$ and 1 null $K$. For each $K$, we calculate the precise positions of three different kinds of spectral switches. 
The results are shown in Fig.~\ref{fig:Figure5}. 
 
The figure shows that the curves are monotonously decreasing with the increase of $K$ from negative to positive. 
This phenomenon can be explained by the decelerating and accelerating property of the curved surfaces. 
The negative curvature decelerate the spectral shift, causing the relative spectral shift curve to be stretched, thereby increasing the distance of spectral switch positions away from the source. 
As $K$ increases, the decelerate property is weakened. 
While the positive curvature accelerates the spectral shift, 
resulting in the relative spectral shift curve shrunk and the positions of spectral switches closer to the source. 
As $K$ increases, the acceleration property becomes more pronounced. 
Therefore, in the process of $K$ changing from negative to positive, 
the spectral switches are first away from the source and then keep approaching the source. 
The curves in the second quadrant are shorter than those in the first quadrant. 
This is because some of the spectral switch points are stretched off the surface when $K<0$. 

By comparing the corresponding curves of different kinds of spectral switches, 
it can also be seen that the sensitivity of different spectral switches to $K$ change is different. 
The overall slope of the curve corresponding to the 2nd spectral switch is greater than that of the 3rd spectral switch, 
and the overall slope of the 3rd is greater than that of the 4th. 
Among these three types of spectral switches, the 2nd spectral switch is the most sensitive, followed by the 3rd and the 4th. 
So the smaller the order of the spectral switch, 
the more sensitive it is to the change of Gaussian curvature $K$. 

\section{Conclusion}
We have investigated the phenomenon of spectral switches in 2D curved space which is described here by surfaces of revolution with constant Gaussian curvature. We have derived the expression of the output on-axis spectrum from a finite source with width $2a$ by using the paraxial approximation. 
By defining the effective propagation distance on curved surfaces, the theoretically approximate solution of the on-axis spectral switch is presented and the on-axis spectral switches are classified. Similar to flat space, spectral switches of fully coherent sources in curved space are also induced by diffraction. 
Nevertheless, the curvature of curved surfaces changes the effective propagation distance of light, 
and thus leads to different effective Fresnel numbers in curved space. This theoretical method can help reduce the amount of calculation of the numerical method and provide the sufficient precision for estimating the positions of spectral switches in curved space with constant Gaussian curvature. 

Comparing the relative spectral shift curves under different $K$, we find that positive Gaussian curvature has a longitudinal compression effect on the relative spectral shift curve, 
and negative Gaussian curvature surface has a longitudinal stretching effect on the curve. 
This effect becomes more and more evident with the increase of $K$, but the jump magnitudes at these discontinuity points on the curves does not change, comparing with that in flat space. 
It also reveals the change of the positions of spectral switches with the change of $K$ for the fully coherent light and shows the influence of the spatial coherence of light on the on-axis spectral change for the polychromatic partially coherent light source during propagation. 
The effects of off-axis spectral switches and various surfaces of revolution are open to be explored further. These results on the spectral effects of light in curved space can promote the understanding about the change of light spectra in non-Euclidean space.  

\acknowledgments{This research is supported by the National Natural Science Foundation of China (NSFC) (grants No.11974309 and 62375241)}.


\begin{thebibliography}{99}
\bibitem{ref1-1-1}G. Rousseaux, C. Mathis, P. Maïssa, T. G. Philbin, and U. Leonhardt, Observation of Negative-Frequency Waves in a Water Tank: A Classical Analogue to the Hawking Effect?, New J. Phys. 10, 053015 (2008).
\bibitem{ref1-1-2}J. Steinhauer, Observation of Quantum Hawking Radiation and Its Entanglement in an Analogue Black Hole, Nature Phys 12, 959 (2016).
\bibitem{ref1-1-3}S. Eckel, A. Kumar, T. Jacobson, I. B. Spielman, and G. K. Campbell, A Rapidly Expanding Bose-Einstein Condensate: An Expanding Universe in the Lab, Phys. Rev. X 8, 021021 (2018).
\bibitem{ref1-1-4}W. G. Unruh and R. Schützhold, On Slow Light as a Black Hole Analogue, Phys. Rev. D 68, 024008 (2003).
\bibitem{ref1-1-5}A. Cortijo and M. A. H. Vozmediano, Electronic Properties of Curved Graphene Sheets, Europhys. Lett. 77, 47002 (2007).
\bibitem{ref1-1}C. Barceló, S. Liberati, and M. Visser, Analogue Gravity, Living Rev. Relativ. 14, 3 (2011).
\bibitem{ref1-2}S. L. Braunstein, M. Faizal, L. M. Krauss, F. Marino, and N. A. Shah, Analogue Simulations of Quantum Gravity with Fluids, Nat Rev Phys (2023).
\bibitem{ref2-1}R. C. T. Da Costa, Quantum Mechanics of a Constrained Particle, Phys. Rev. A 23, 1982 (1981).
\bibitem{ref2-2}S. Longhi, Topological Optical Bloch Oscillations in a Deformed Slab Waveguide, Opt. Lett. 32, 2647 (2007).
\bibitem{ref2-3}G. Della Valle, M. Savoini, M. Ornigotti, P. Laporta, V. Foglietti, M. Finazzi, L. Duò, and S. Longhi, Experimental Observation of a Photon Bouncing Ball, Phys. Rev. Lett. 102, 180402 (2009).
\bibitem{ref2-4}A. Szameit, F. Dreisow, M. Heinrich, R. Keil, S. Nolte, A. Tünnermann, and S. Longhi, Geometric Potential and Transport in Photonic Topological Crystals, Phys. Rev. Lett. 104, 150403 (2010).
\bibitem{ref2-5}A. Libster-Hershko, R. Shiloh, and A. Arie, Surface Plasmon Polaritons on Curved Surfaces, Optica 6, 115 (2019).
\bibitem{ref2-5-1}J. Bělín, T. Tyc, and S. A. R. Horsley, Optical Simulation of Quantum Mechanics on the Möbius Strip, Klein’s Bottle and Other Manifolds, and Talbot Effect, New J. Phys. 23, 033003 (2021).
\bibitem{ref2-6}S. Batz and U. Peschel, Linear and Nonlinear Optics in Curved Space, Phys. Rev. A 78, 043821 (2008).
\bibitem{ref2-7}V. H. Schultheiss, S. Batz, A. Szameit, F. Dreisow, S. Nolte, A. Tünnermann, S. Longhi, and U. Peschel, Optics in Curved Space, Phys. Rev. Lett. 105, 143901 (2010).
\bibitem{ref2-8}V. H. Schultheiss, S. Batz, and U. Peschel, Hanbury Brown and Twiss Measurements in Curved Space, Nature Photon 10, 106 (2016).
\bibitem{ref2-9}C. Xu, A. Abbas, L.-G. Wang, S.-Y. Zhu, and M. S. Zubairy, Wolf Effect of Partially Coherent Light Fields in Two-Dimensional Curved Space, Phys. Rev. A 97, 063827 (2018).
\bibitem{ref2-10}A. Patsyk, M. A. Bandres, R. Bekenstein, and M. Segev, Observation of Accelerating Wave Packets in Curved Space, Phys. Rev. X 8, 011001 (2018).
\bibitem{ref2-11}A. Patsyk, U. Sivan, M. Segev, and M. A. Bandres, Observation of Branched Flow of Light, Nature 583, 60 (2020).
\bibitem{ref2-12}R. Bekenstein, Y. Kabessa, Y. Sharabi, O. Tal, N. Engheta, G. Eisenstein, A. J. Agranat, and M. Segev, Control of Light by Curved Space in Nanophotonic Structures, Nature Photon 11, 664 (2017).
\bibitem{ref2-13}C. Xu, A. Abbas, and L.-G. Wang, Generalization of Wolf Effect of Light on Arbitrary Two-Dimensional Surface of Revolution, Opt. Express 26, 33263 (2018).
\bibitem{ref2-14}C. Xu and L.-G. Wang, Theory of Light Propagation in Arbitrary Two-Dimensional Curved Space, Photon. Res. 9, 2486 (2021).
\bibitem{ref3-1}B. A. Garetz, Angular Doppler Effect, J. Opt. Soc. Am. 71, 609 (1981).
\bibitem{ref3-2}M. P. J. Lavery, F. C. Speirits, S. M. Barnett, and M. J. Padgett, Detection of a Spinning Object Using Light’s Orbital Angular Momentum, Science 341, 537 (2013).
\bibitem{ref3-3}E. Wolf, Invariance of the Spectrum of Light on Propagation, Phys. Rev. Lett. 56, 1370 (1986).
\bibitem{ref3-4}E. Wolf, Non-Cosmological Redshifts of Spectral Lines, Nature 326, 363 (1987).
\bibitem{ref3-5}G. M. Morris and D. Faklis, Effects of Source Correlation on the Spectrum of Light, Optics Communications 62, 5 (1987).
\bibitem{ref3-6}M. F. Bocko, D. H. Douglass, and R. S. Knox, Observation of Frequency Shifts of Spectral Lines Due to Source Correlations, Phys. Rev. Lett. 58, 2649 (1987).
\bibitem{ref3-7}E. Wolf, T. Shirai, H. Chen, and W. Wangh, Coherence Filters and Their Uses. I. Basic Theory and Examples, Journal of Modern Optics 44, 1345 (1997).
\bibitem{ref3-8}T. Shirai, E. Wolf, H. Chen, and W. Wang, Coherence Filters and Their Uses II. One-Dimensional Realizations, Journal of Modern Optics 45, 799 (1998).
\bibitem{ref4-1}J. Pu, H. Zhang, and S. Nemoto, Spectral Shifts and Spectral Switches of Partially Coherent Light Passing through an Aperture, Optics Communications 162, 57 (1999).
\bibitem{ref4-2}J. T. Foley and E. Wolf, Phenomenon of Spectral Switches as a New Effect in Singular Optics with Polychromatic Light, J. Opt. Soc. Am. A 19, 2510 (2002).
\bibitem{ref4-3}J. Pu, S. Nemoto, and B. Lü, Effect of Spectral Correlations on Spectral Switches in the Diffraction of Partially Coherent Light, J. Opt. Soc. Am. A 20, 1933 (2003).
\bibitem{ref4-4}H. C. Kandpal, Experimental Observation of the Phenomenon of Spectral Switch, J. Opt. A: Pure Appl. Opt. 3, 296 (2001).
\bibitem{ref4-5}H. C. Kandpal, S. Anand, and J. S. Vaishya, Experimental Observation of the Phenomenon of Spectral Switching for a Class of Partially Coherent Light, IEEE J. Quantum Electron. 38, 336 (2002).
\bibitem{ref4-6}S. Anand, B. K. Yadav, and H. C. Kandpal, Experimental Study of the Phenomenon of 1 × N Spectral Switch Due to Diffraction of Partially Coherent Light, J. Opt. Soc. Am. A 19, 2223 (2002).
\bibitem{ref4-7}B. K. Yadav, S. A. M. Rizvi, and H. C. Kandpal, Experimental Observation of Spectral Changes of Partially Coherent Light in Young’s Experiment, J. Opt. A: Pure Appl. Opt. 8, 72 (2006).
\bibitem{ref4-8}J. Li and L. Chang, Spectral Shifts and Spectral Switches of Light Generated by Scattering of Arbitrary Coherent Waves from a Quasi-Homogeneous Media, Opt. Express 23, 16602 (2015).
\bibitem{ref4-9}X. Wang, Z. Liu, and K. Huang, Multiple Spectral Switches Generated by the Scattering of a Novel Electromagnetic Random Source of Circular Frame upon a Semisoft Boundary Medium, Laser Phys. Lett. 16, 066005 (2019).
\bibitem{ref4-10}Y. Zhang and J. Zhou, Spectral Shifts and Spectral Switches of Polychromatic Stochastic Electromagnetic Vortex Beams on Scattering from a Semisoft Boundary Medium, J. Opt. 21, 045608 (2019).
\bibitem{ref4-11}P. Han, Near Field Lattice Spectroscopy with a Reflective Confocal Configuration, Appl. Phys. Express 4, 022401 (2011).
\bibitem{ref4-12}E. Wolf and D. F. V. James, Correlation-Induced Spectral Changes, Rep. Prog. Phys. 59, 771 (1996).
\bibitem{ref4-13}J. Pu, C. Chi, and S. Nemoto, Spectral Anomalies in Young’s Double-Slit Interference Experiment, Opt. Express 12, 5131 (2004).
\bibitem{ref4-14}P. Han, All Optical Spectral Switches, Opt. Lett. 37, 2319 (2012).
\bibitem{MandelWolf1995}L. Mandel and E. Wolf, Optical Coherence and Quantum Optics, 1st ed. (Cambridge University Press, 1995).
\end{thebibliography}
\end{document}